\documentclass[aps,twocolumn,amsmath,amssymb, superscriptaddress]{revtex4}
\usepackage[pdftex]{graphicx}% Include figure files
 \usepackage{amsmath}
\usepackage[utf8x]{inputenc}
\usepackage[T1]{fontenc}
\usepackage{subfigure}
\usepackage{amssymb}
\usepackage{flushend}
 \usepackage{tikz}
\usetikzlibrary{decorations.markings}
\usepackage{amsfonts}
\usepackage{bm}
 \usepackage{amsmath} 
\usepackage{amsfonts} 
\usepackage{amssymb, mathrsfs}
\usepackage{braket}
\usepackage{graphicx} 

\usepackage{bbm}
\def\beq{\begin{equation}}
\def\eeq{\end{equation}}
\def\bsp{\begin{split}}
\def\esp{\end{split}}
\def\bea{\begin{eqnarray}}
\def\eea{\end{eqnarray}}
\def\ba{\begin{array}}
\def\ea{\end{array}}

\def\dg{\dagger}

\def\lb{\left(}
\def\rb{\right)}

\def\l.{\left.}
\def\r.{\right.}

\def\ra{\rangle}
\def\la{\langle}

\def\bo{\bold{k}}

\def\dg{\dagger}

\begin{document}
\author{S. A. Owerre}

\title{Magnon Hall Effect without Dzyaloshinskii-Moriya Interaction}
\affiliation{ Perimeter Institute for Theoretical Physics, 31 Caroline St. N., Waterloo, Ontario N2L 2Y5, Canada.}
\affiliation{ African Institute for Mathematical Sciences, 6 Melrose Road, Muizenberg, Cape Town 7945, South Africa.}
%\email{solomon@aims.ac.za}

\begin{abstract}
Topological magnon bands and magnon Hall effect in insulating collinear ferromagnets  are induced by the Dzyaloshinskii-Moriya interaction (DMI) even at zero magnetic field. In the geometrically frustrated star lattice, a coplanar/noncollinear $\bold q=0$ magnetic ordering may be present due to spin frustration. This magnetic structure, however,  does not exhibit  topological magnon effects  even with DMI in contrast to collinear  ferromagnets. We show that a magnetic field applied perpendicular to the star plane induces a non-coplanar spin configuration with nonzero spin scalar chirality, which provides topological effects without the need of DMI. The non-coplanar spin texture originates from the topology of the spin configurations and does not need the presence of DMI or magnetic ordering, which suggests that this phenomenon  may be present in the chiral spin liquid phases of frustrated magnetic systems.    We propose that these anomalous topological magnon effects can be accessible in Polymeric Iron (III) Acetate --- a star-lattice antiferromagnet with both spin frustration and long-range magnetic ordering.

\end{abstract}

\maketitle
Recently, the experimental observation of thermal Hall effect of spin excitations has been reported in the frustrated  Kagom\'e  volborthite  Cu$_3$V$_2$O$_7$(OH)$_2$$\cdot$2H$_2$O \cite{wat} and frustrated honeycomb  antiferromagnet Ba$_3$CuSb$_2$O$_9$ \cite{wat0}, with no signs of DMI. This effect has been previously observed in collinear ferromagnetic materials with DMI \cite{alex6,alex1,alex1a} and  pyrochlore spin liquid material \cite{hir}. In these recent reports, a transverse thermal Hall conductivity $\kappa_{xy}$  was observed in a strong magnetic field  $\sim 15~\text{Tesla}$ applied perpendicular to the plane of the frustrated magnets \cite{wat0, wat}. The observed effect on the Kagom\'e  volborthite is attributed to spin excitations in the spin liquid (SL) regime. However, the Kagom\'e  volborthite is known to exhibit different magnetic-field-induced ordered phases for magnetic fields $< 15 ~\text{Tesla}$ \cite{Yo,Yo1}. The frustrated Kagom\'e compound Ca$_{10}$Cr$_7$O$_{28}$ \cite{Balz} also exhibits ferromagnetic ordered states for magnetic field of magnitude  $\sim 11 ~\text{Tesla}$. This suggests that the observed low temperature dependence of $\kappa_{xy}$ in Kagom\'e  volborthite might not be due to spin excitations in the SL regime, but magnon excitations in the field-induced ordered phases. Following these recent developments, we have recently shown \cite{owe} that the profile of $\kappa_{xy}$ in Kagom\'e  volborthite  can be captured quantitatively by considering the topological magnon bands in the Kagom\'e antiferromagnet with/without DMI \cite{dm}. 

The honeycomb-lattice antiferromagnet Ba$_3$CuSb$_2$O$_9$ also shows a negative $\kappa_{xy}$ at the same magnetic field and a power-law temperature dependence $\kappa_{xy}\propto T^2$ \cite{wat0}. It was suggested that the observed thermal Hall effect is a phonon Hall effect. However, a closely related  honeycomb-lattice antiferromagnetic material Bi$_3$Mn$_4$O$_{12}$(NO$_3$) \cite{matt1} shows evidence of magnetic order at a critical field of $\sim 6~\text{Tesla}$ consistent with a collinear N\'eel order \cite{matt}. Prior to this experimental report,  we have already shown that honeycomb (anti)ferromagnet  with a next-nearest-neighbour staggered DMI  captures a negative $\kappa_{xy}$ and power-law temperature dependence $\kappa_{xy}\propto T^2$ \cite{owe0, owe1} as recently seen in Ba$_3$CuSb$_2$O$_9$  \cite{wat0}. In this regard, we believe that this correspondence between theory and experiment cannot be serendipitous. There must be an evidence of field-induced magnetic order in these frustrated antiferromagnetic materials  and the associated $\kappa_{xy}$ must be related to that of magnon excitations.

The star-lattice antiferromagnet is definitely  another interesting candidate for realizing nontrivial excitations and thermal Hall conductivity. This lattice can be considered  as a variant of the Kagom\'e lattice by introducing additional lattice links between triangles of the Kagom\'e lattice. It is also closely related to the honeycomb lattice by shrinking the three-site  triangles as one site. However, the star-lattice contains six sites in the unit cell as opposed to the Kagom\'e and honeycomb lattices. In fact, many different models show interesting features on this lattice  \cite{zheng1,ric, zheng, zheng3, zheng4,zheng2, zheng5, zheng6}. A common known material with this lattice structure is Polymeric Iron(III) Acetate, Fe$_3$($\mu_3$-O)($\mu$-OAc)$_6$(H$_2$O)$_3$[Fe$_3$($\mu_3$-O)($\mu$-OAc)$_{7.5}$]$_2\cdot$ 7H$_2$O, which  carries a spin moment of $S=5/2$. In this material, both spin frustration and long-range magnetic ordering coexist at low temperatures \cite{zheng}, but a magnetic field is sufficient to circumvent the spin frustrations and pave the way for long-range magnetic ordering with magnon excitations. 

In this Letter, we study the topological properties of geometrically frustrated star lattice antiferromagnet. We focus on the coplanar/noncollinear $\bold q=0$ N\'eel state, which is definitely a long-range magnetic ordering on the star-lattice induced by spin frustration. In the absence of both the magnetic field and the DMI, there are two flat modes consisting of one zero mode,  and four dispersive modes. A nearest-neighbour DMI is known to stabilize the $\bold q=0$ N\'eel state on the Kagom\'e lattice \cite{men0, men1, men2, men3, men4, men4a,zor}. This is likely the case on the  star lattice. However, in stark contrast to ferromagnets \cite{kk1,kk2,kk3,kk4,kk5,kk6,kk7, shin1},  the DMI does not lead to topological properties  in the $\bold q=0$ N\'eel state. Hence, the magnon bands remain gapless,  leading to  vanishing $\kappa_{xy}$ and  no  protected chiral edge states. In the presence of an out-of-plane magnetic field, the coplanar $\bold q=0$ N\'eel state becomes a non-collinear/non-coplanar spin texture.  We show that topological effects are induced by the magnetic field via an induced chiral interaction $H_{\chi}\sim\cos\chi\sum \bold S_{i}\cdot \lb \bold S_{j}\times\bold S_{k}\rb$, where $\cos\chi\propto \text{magnetic field}$.    The resulting  magnon bands are gapped. We observe a finite $\kappa_{xy}$ and protected magnon edge states, which persist for zero DMI \cite{footnote}. It is important to note that  the spin scalar chirality  survives even in the absence of magnetic ordering $\la {\bf S}_j\ra=0$, therefore topological effects may be present in chiral spin liquid phase of the star lattice.  The proposed phenomenon is very likely to occur in Polymeric Iron(III) Acetate \cite{zheng}.

\begin{figure}
\centering
\includegraphics[width=.7\linewidth]{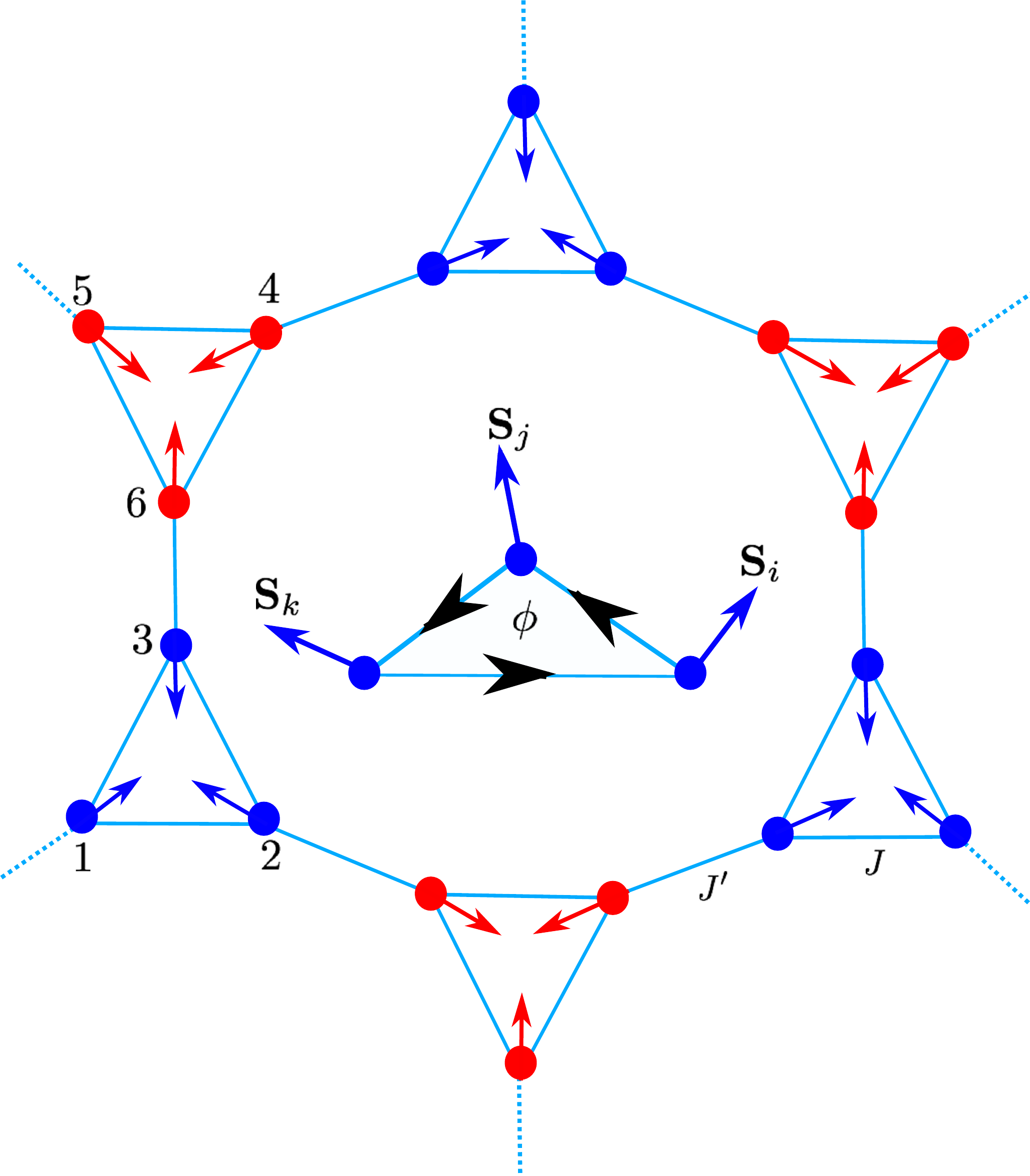}
\caption{Color online. The zero field  coplanar $\bold q=0$ N\'eel order on the geometrically frustrated star-lattice.  The numbers  denote different sublattices. Inset: A nonzero out-of-plane magnetic field  generates a non-coplanar spin texture with field-induced fictitious flux ($\phi$) within each triangular plaquette.   In Polymeric Iron(III) Acetate \cite{zheng}  $J^\prime>J$,  hence  $J^\prime/J>1$. }
\label{fig1}
\end{figure}

 The  model Hamiltonian  for our study is given by
\begin{align}
H&= \sum_{\la i, j\ra} J_{ij} {\bf S}_{i}\cdot{\bf S}_{j}+\sum_{\la i, j\ra} \bold{D}_{ij}\cdot{\bf S}_{i}\times{\bf S}_{j}-h\hat{\bold z}\cdot\sum_i {\bf S}_i,
\label{apen1}
\end{align}
where $J_{ij}=J,J^\prime>0$ are isotropic antiferromagnetic couplings within and between triangles as shown in Fig.~\ref{fig1}. $\bold D_{ij}$ is the DMI between sites $i$ and $j$ within triangles, and $h$ is the magnitude of the out-of-plane magnetic field in units of $g\mu_B$.  On the Kagom\'e lattice, the $\bold q=0$ ground state is known to be stabilized  by an antiferromagnetic  next-nearest-neighbour exchange \cite{harr} or an out-of-plane DMI, $\bold D_{ij}=(0,0, \mp D_z)$ \cite{men1}, where $\mp$ alternates between down and up pointing triangles respectively. Although the DMI alternates between the triangles, only one ground state is selected for each sign with $D_z>0$ (positive chirality) and $D_z<0$ (negative chirality).  In principle a DMI is present on the star-lattice since the midpoint between the bonds connecting two sites is not a center of inversion similar to the Kagom\'e lattice. Hence,  we will assume that the  out-of-plane DMI stabilizes  the $\bold q=0$ ground state on the star-lattice. It is important to note that the $\bold q=0$ ground state can equally be stabilized through other anisotropy interactions \cite{men4a}. In most materials, an in-plane DMI may be present, however this component does not induce any topological magnon bands (see Ref.~\cite{men4a}) and it is usually small and can be neglected for simplicity as it does not change any results of this Letter.

\begin{figure}
\centering
\includegraphics[width=1\linewidth]{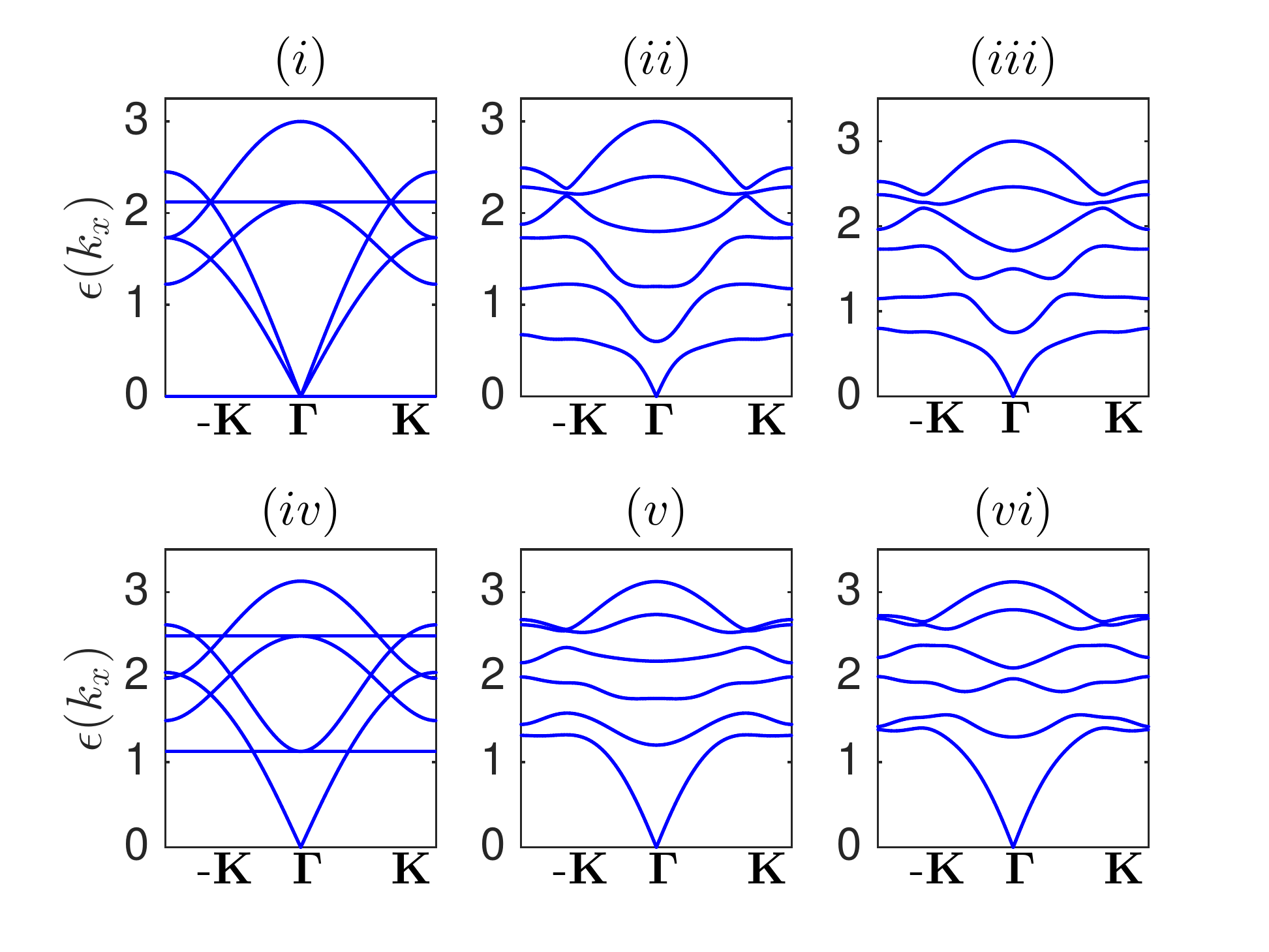}
\caption{Color online. Magnon band structure along $k_y=0$ with $J^\prime/J=1.5$. For the upper panel $D_z/J=0$:  $(i)~h=0,~(ii)~h/h_s=0.2,~(iii)~h/h_s=0.25$. For the upper panel $D_z/J=0.15$: $(iv)~h=0,~(v)~h/h_s=0.2,~(vi)~h/h_s=0.25$. The linear gapless dispersion of the lowest band at ${\bf \Gamma}=(0,0)$ signifies antiferromagnetic order.}
\label{m_bands}
\end{figure}
\begin{figure}
\centering
\includegraphics[width=1\linewidth]{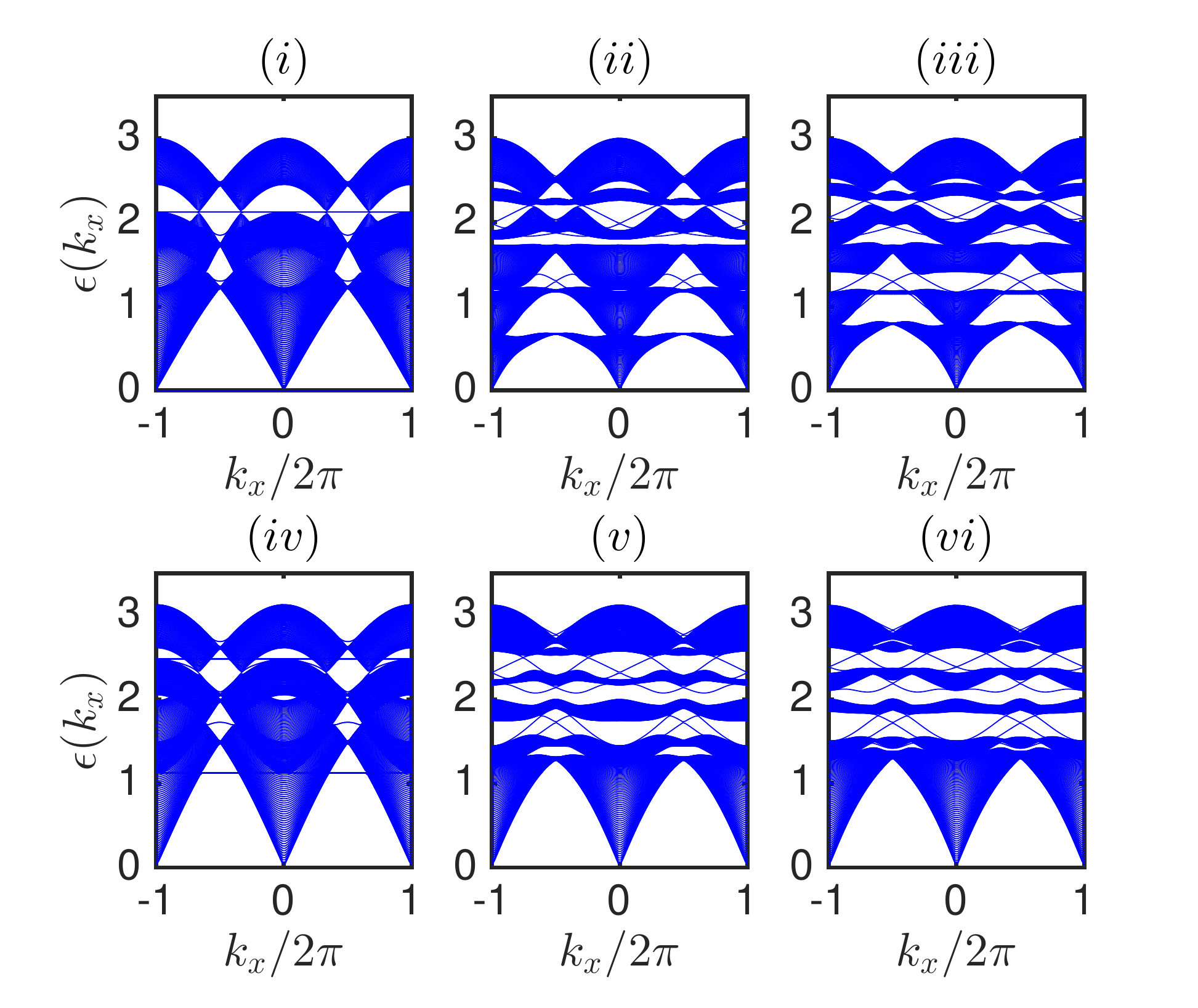}
\caption{Color online. The corresponding magnon edge states of Fig.~\ref{m_bands} for a strip geometry.}
\label{edge}
\end{figure}

 In the classical limit, the spin operators  can be approximated as classical vectors,  written as
 $\bold{S}_{i}= S\bold{n}_i$, where $\bold{n}_i=\lb\sin\chi\cos\theta_i, \sin\chi\sin\theta_i,\cos\chi \rb$
 is a unit vector and $\theta_i$ labels the spin oriented angles on each sublattice and $\chi$ is the field-induced canting angle. For the $\bold q=0$ N\'eel order in Fig.~\ref{fig1},  $\theta_1=5\pi/3,~\theta_2=\pi/3,~\theta_3=\pi,~\theta_4=2\pi/3,~\theta_5=4\pi/3,~\theta_6=0$. The classical energy is given by
\begin{align}
e_0&= -\frac{J}{2}\lb 1 - {3}\cos^2\chi\rb-\frac{J^\prime}{2}\lb 1 - 2\cos^2\chi\rb \\&\nonumber-\frac{\sqrt{3}}{2}D_z\sin^2\chi-h\cos\chi,
\end{align}
where $e_0=E_{cl}/6NS^2$ and  the magnetic field is rescaled in unit of $S$. Minimizing this energy yields the canting angle $\cos\chi = h/h_s$, where $h_s=(3J+2J^\prime+\sqrt{3}D_z)$ is the saturation field. For the excitations above the classical ground state, the general procedure is as follows. At zero field, the spins lie on the plane of the star lattice taken as the $x$-$y$ plane as shown in Fig.~\ref{fig1}. Then, we perform a rotation  about the $z$-axis on the sublattices by the spin oriented  angles in order to achieve the 120$^\circ$ coplanar  N\'eel order. At this point the quantization axis can be chosen as the $y$-axis. As the out-of-plane magnetic field is turned on, the spins  cant towards the direction of the field and form a non-coplanar configuration (see inset of Fig.~\ref{fig1}). Thus, we have to align them along the new quantization axis by performing a rotation about the $y$-axis by the field canting angle $\chi$.  Hence, \bea \bold{S}_i=\mathcal{R}_z(\theta_i)\cdot\mathcal{R}_y(\chi)\cdot\bold S_i^\prime,\eea
where
\begin{align}
\mathcal{R}_z(\theta_i)\cdot\mathcal{R}_y(\chi)
=\begin{pmatrix}
\cos\theta_i\cos\chi & -\sin\theta_i & \cos\theta_i\sin\chi\\
\sin\theta_i\cos\chi & \cos\theta_i &\sin\theta_i\sin\chi\\
-\sin\chi & 0 &\cos\chi
\end{pmatrix}.
\end{align}
    We consider the positive chirality ground states,   $\bold{D}_{ij}=(0,0,-D_z)$ with $D_z>0$. The corresponding Hamiltonian that contribute to noninteracting magnon model  is given by
  \begin{align}
  H_{J}&= J\sum_{\la i, j\ra}\bigg[\cos\theta_{ij} \bold S_{i}^\prime\cdot \bold S_{j}^\prime+ \sin\theta_{ij}\cos\chi \hat{\bold z}\cdot\lb \bold S_{i}^\prime\times\bold S_{j}^\prime\rb  \\&\nonumber +2\sin^2\lb\frac{\theta_{ij}}{2}\rb[\sin^2\chi S_{i}^{\prime x}S_{j}^{\prime x} +\cos^2\chi S_{i}^{\prime z}S_{j}^{\prime z}]\bigg] , \\
    H_{J^\prime}&= J^\prime\sum_{ \la i, j\ra}\bigg[\cos\theta_{ij} \bold S_{i}^\prime\cdot \bold S_{j}^\prime+ \sin\theta_{ij}\cos\chi \hat{\bold z}\cdot\lb \bold S_{i}^\prime\times\bold S_{j}^\prime\rb  \\&\nonumber +2\sin^2\lb\frac{\theta_{ij}}{2}\rb[\sin^2\chi S_{i}^{\prime x}S_{j}^{\prime x} +\cos^2\chi S_{i}^{\prime z}S_{j}^{\prime z}]\bigg] , \\
   H_{DMI}&= D_z\sum_{\la i, j\ra}\bigg[ \sin\theta_{ij}[\cos^2\chi S_{i}^{\prime x}S_{j}^{\prime x} + S_{i}^{\prime y}S_{j}^{\prime y}\\&\nonumber+\sin^2\chi S_{i}^{\prime z}S_{j}^{\prime z}]-\cos\theta_{ij}\cos\chi \hat{\bold z}\cdot\lb \bold S_{i}^\prime\times\bold S_{j}^\prime\rb\bigg], \\
   H_z&=-h\cos\chi\sum_{i} S_{i}^{\prime z},
  \end{align}
  where $\theta_{ij}=\theta_i-\theta_j$. The spin scalar chirality $\bold S_k^\prime\cdot\lb \bold S_{i}^\prime\times\bold S_{j}^\prime\rb$ with $\bold S_k^\prime=\hat{\bold z}$ is generated by the magnetic field applied perpendicular to the star plane as depicted in the inset of Fig.~\ref{fig1}.  Evidently, the scalar chirality survives at $D_z=0$, therefore the presence of the DMI is not necessarily needed provided the $\bold q=0$ ordering is stabilized \cite{footnote}. As mentioned previously,  spin scalar chirality is nonzero even when there is no magnetic ordering $\la {\bf S}_j\ra=0$ and can be used as the order parameter in geometrically frustrated systems such as chiral spin liquid states. Therefore, it is possible that the basic result of this Letter can be observed in systems without magnetic ordering but exhibits non-coplanar spin configuration.      This is the major difference between the present model and previously studied collinear ferromagnets \cite{kk1,kk2,kk3,kk4,kk5,kk6,kk7, shin1}.

\begin{figure}
\centering
\includegraphics[width=.9\linewidth]{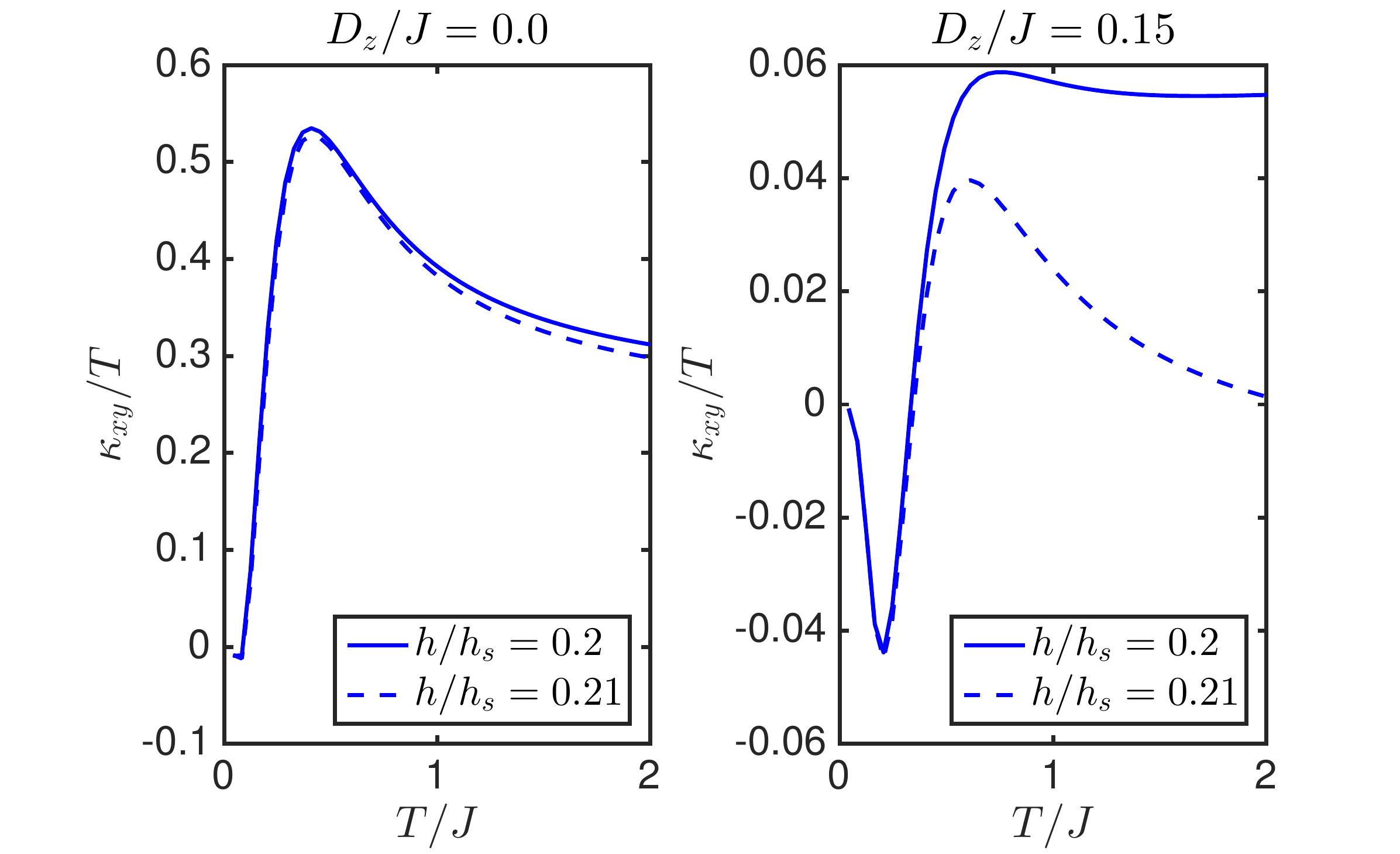}
\caption{Color online.   Low-temperature dependence of $\kappa_{xy}$  for two separate DMI  and field values at $J^\prime/J=1.5$.}
\label{BB1}
\end{figure}

Interestingly, the coefficient of the chiral interaction $H_{\chi}$ vanishes for the $J^\prime$ coupling between triangles since  $\sin\theta_{ij}=0$ in the $J^\prime$ term. This is consistent with the triangular geometry of the star-lattice.  Now, we   express the spin operators in terms of the Holstein-Primakoff spin boson operators \cite{HP} $S_{i}^{\prime x}=\sqrt{S/2}(b_{i\mu}^\dg+b_{i\mu})$, $S_{i}^{\prime y}=i\sqrt{S/2}(b_{i\mu}^\dg-b_{i\mu})$ and $S_{i}^{\prime z}=S-b_{i,\mu}^\dg b_{i,\mu}$.  In momentum space the Hamiltonian can be written as \bea H=\frac{S}{2}\sum_{\bold k}\psi^\dagger_{\bold k}\cdot \boldsymbol{\mathcal{H}}(\bold k)\cdot\psi_{\bold k},\eea with $\psi^\dagger_{\bold k}= (b_{\mu,\bold{k}}^{\dagger},\thinspace b_{\mu^\prime,\bold{k}}^{\dagger}, b_{\mu,-\bold{k}},\thinspace b_{\mu^\prime,-\bold{k}})$, where $\mu=1,2,3$ and  $\mu^\prime=4,5,6$. The Bogoliubov Hamiltonian  $\boldsymbol{\mathcal{H}}(\bold k)$ is a $12\times 12$ matrix given by

\begin{align}
\boldsymbol{\mathcal{H}}(\bold k)= 
\begin{pmatrix}
\bold{A}(\bo, \phi) & \bold{B}(\bo)\\
\bold{B}^*(-\bo)&\bold{A}^*(-\bo,\phi)
\end{pmatrix},
\label{ham}
\end{align}
where  \begin{align}
\bold{A}(\bold k)= 
\begin{pmatrix}
\bold{a}_1(\phi) & \bold{b}_1(\bo)\\
\bold{b}_1(-\bo)&\bold{a}_1(\phi)
\end{pmatrix},~\bold{B}(\bold k)= 
\begin{pmatrix}
\bold{a}_2 & \bold{b}_2(\bo)\\
\bold{b}_2(-\bo)& \bold{a}_2
\end{pmatrix},
\end{align}
 
\begin{align}
&\bold{a}_1(\phi)= 
\begin{pmatrix}
\Delta_0&\Delta e^{-i\phi}&\Delta e^{i\phi}\\
\Delta e^{i\phi}&\Delta_0&\Delta e^{-i\phi}\\
\Delta e^{-i\phi}&\Delta e^{i\phi}&\Delta_0
\end{pmatrix},~
\bold{a}_2= \Delta^\prime 
\begin{pmatrix}
0&1&1\\
1&0&1\\
1&1&0
\end{pmatrix}\\&
\bold{b}_1(\bo)=\Lambda
\begin{pmatrix}
 e^{ik_2}&0&0\\
0& e^{ik_1}&0\\
0&0&1
\end{pmatrix},~
\bold{b}_2(\bo)=\Lambda^\prime
\begin{pmatrix}
e^{ik_2}&0&0\\
0& e^{ik_1}&0\\
0&0&1\\
\end{pmatrix},
\end{align}
where $\Delta_0=h_{\chi}-(\Delta_z + \Lambda_z)=J+J^\prime +\sqrt{3}D_z$, $k_{1}=\bold{k}\cdot \bold{e}_{1}$ and $k_{2}=\bold{k}\cdot \bold{e}_{2}$. The lattice basis vectors are chosen as $\bold e_1=2\hat{\bold x}$ and  $\bold e_2=\hat{\bold x} + \sqrt{3}\hat{\bold y}$. The coefficients are given by
\begin{align}
&\Delta_z= 2\bigg[J\lb -\frac{1}{2}+\frac{3}{2}\cos^2\chi\rb-\frac{ \sqrt{3}D_z}{2}\sin^2\chi\bigg]\\
&\Delta=\sqrt{(\Delta_R)^2+(\Delta_M)^2},\\
&\Delta_R= J\lb-\frac{1}{2} +\frac{3\sin^2\chi}{4}\rb-\frac{\sqrt{3}D_z}{2}\lb 1-\frac{\sin^2\chi}{2}\rb,\\
&\Delta_M= \cos\chi\lb -\frac{\sqrt{3}J}{2}+\frac{D_z}{2}\rb,\\
&\Delta^\prime=\frac{\sin^2\chi}{2}\lb \frac{3J}{2}+\frac{\sqrt{3}D_z}{2}\rb,\\
&\Lambda_z= J^\prime\lb -1+2\cos^2\chi\rb,~
\Lambda= J^\prime\lb-1 +\sin^2\chi\rb,\\
& \Lambda^\prime=J^\prime\sin^2\chi,~h_\chi=h\cos\chi,
\end{align}
and $\tan\phi_{ij}=\Delta_M/\Delta_R$.   Notice that the fictitious magnetic flux does not vanish at zero DMI unlike in ferromagnets. 

In Polymeric Iron(III) Acetate \cite{zheng}, the intra-layer coupling $J$ is weaker than the inter-layer coupling $J^\prime$, hence $J^\prime/J>1$.  We have shown the magnon bands in Fig.~\ref{m_bands} for $J^\prime/J=1.5$: $D_z/J=0$ (upper panel) and $D_z/J=0.15$ (lower panel) with several values of the magnetic field $h/h_s$ along the Brillouin zone (BZ) line $\pm {\bf K}=(\pm 2\pi/3, 0)$. For  $D_z/J=0$, the system exhibits  two flat modes and four dispersive bands at $h/h_s=0$.  The flat modes contain one zero mode due to the geometry of the star-lattice, and the magnon bands are completely gapless at various points in the BZ. At zero DMI $D_z/J=0$, a moderate increase in the magnetic field    lifts the flat zero mode and induces gaps at various points in the magnon bands. Notice that the flat modes also acquire a small dispersion.

 For $D_z/J=0.15$ the zero mode is lifted  at $h/h_s=0$, but the magnon bands remain gapless. As mentioned above, this is due to the fact that the presence of the DMI does not have any topological effects on the $\bold q=0$ N\'eel state.  In fact, this is the major difference between the present model and previously studied collinear ferromagnets \cite{kk1,kk2,kk3,kk4,kk5,kk6,kk7,shin1}. As the magnetic field increases from zero the flat modes  acquire a small dispersion and the magnon bands also acquire a gap similar to the case without DMI. The linear gapless dispersion of the lowest band at ${\bf \Gamma}$ signifies antiferromagnetic order.
 
 In order to substantiate the nontrivial topology of this system at finite magnetic field,  we have solved for the  magnon edge states for a strip geometry on the star-lattice as shown in Fig.~\ref{edge}. For zero magnetic field, there is no counter-propagating gapless edge states and  the Chern number is zero for all bands, confirming the fact that the system is topologically trivial at zero field. In contrast,  for finite magnetic field  counter-propagating gapless edge states are discernible in Fig.~\ref{edge} with a Chern number of $\pm 1$ signifying the strong topology of the system for nonzero magnetic field irrespective of the DMI. Furthermore, we have confirmed the strong topology of this system at finite field by computing the transverse thermal Hall conductivity $\kappa_{xy}$ \cite{kk2,shin1}. Figure~\ref{BB1} shows the low-temperature dependence of $\kappa_{xy}$ for $D_z/J=0$ and $D_z/J=0.15$ with several field values. As expected, $\kappa_{xy}$ vanishes at zero magnetic field, and a non-vanishing $\kappa_{xy}$ is present at finite magnetic field and persists for zero DMI \cite{footnote}.

In summary, the results of this Letter is not simply a consequence of time-reversal symmetry (TRS) breaking, because  the magnetic order that underlies magnons has already broken TRS even in ferromagnets. Nevertheless, topological effects do not emerge in the conventional magnonic systems even though TRS is already broken. Another feature of this model is that the magnon bands are not doubly degenerate at zero field as one would expect in TRS invariant systems. This is because  magnons are bosonic quasiparticles and the TR operator is defined as $\mathcal T^2=+1$, which does not obey Kramers theorem.   In  this model the  broken inversion symmetry of the lattice allows a DMI, but its role  is different from ferromagnets, since  the coplanar/noncollinear $\bold q=0$ spin configuration  is a consequence of geometric frustration. The basic result of this Letter is that this magnetic ordering is not topological and we showed that  topological effects require a topological non-coplanar spin texture with a finite spin scalar chirality. This result originates from the topology of the spin configuration without the need of DMI. It also means that any spin configuration with a non-coplanar structure will exhibit the same effect even when they are not necessarily ordered. Topological Hall effect in non-coplanar systems has been observed in various  frustrated electron systems \cite{ele0,ele1,ele2,ele}. The present model is a magnonic system and we believe that these results  can be accessible experimentally in the present and upcoming star-lattice quantum magnetic materials, and can be probed by using neutron inelastic scattering.   The magnon edge modes can be probed by edge sensitive methods such as  light \cite{luuk} or electronic \cite{kha} scattering method.  The experimental study of topological magnon bands and edge state modes are the subjects of current interest \cite{rc}.

Research at Perimeter Institute is supported by the Government of Canada through Industry Canada and by the Province of Ontario through the Ministry of Research
and Innovation.


\begin{thebibliography}{0}
\bibitem{footnote} 
 For zero DMI, the  coplanar $\bold q=0$ magnetic order can be stabilized by other non-chiral anisotropy  interactions such as an easy plane  anisotropy. In this case, the magnetic field still induces a non-coplanar spin texture with nonzero spin scalar chirality and a nonzero thermal Hall conductivity as shown in the text.
\end{thebibliography}
\end{document}